\documentclass[jkps,preprint,fleqn,showpacs,showkeys]{revtex4}
\usepackage{graphicx}
\usepackage{amssymb}
\usepackage{bm}
\begin{document}
\title{CP Violation in the Neutral Higgs Sector of a Non-minimal Supersymmetric Standard Model
with Multiple Higgs Singlets}

\author{S. W. \surname{Ham}}
\email{s.w.ham@hotmail.com}
\affiliation{Department of Physics, Korea Advanced Insititute of Science and Technology, Daejon 305-701}

\author{S. K. \surname{Oh}}
\affiliation{Department of Physics, Konkuk University, Seoul 143-701,
and Center for High Energy Physics, Kyungpook National University,
Daegu 702-701 }

\begin{abstract}
The possibility of CP violation is studied in the Higgs sector of a supersymmetric standard model
with multiple Higgs singlets.
The tree-level Lagrangian in this model is assumed to conserve the CP symmetry.
We find that CP violation is viable in this model at the one-loop level, in an explicit way,
if the radiative corrections from the third generation of quarks and squarks are taken into account.
In the presence of explicit CP violation, at the one-loop level,
the upper bound on the mass of the lightest neutral Higgs boson
and the productions of the neutral Higgs bosons via the Higgsstrahlung process
in high-energy $e^+e^-$ collisions are calculated.
We find that the upper bound on the mass of the lightest neutral Higgs boson increases
as the number of Higgs singlets increases in a regulated manner.
The production cross sections of the neutral Higgs bosons also show a reasonable increasing behavior
with respect to the number of Higgs singlets.
\end{abstract}

\maketitle
\section{Introduction}

The Standard Model (SM) possesses a complex phase in the Cabbibo Kobayashi Maskawa (CKM) matrix,
but the size of the complex phase in the SM, as determined by several precision experiment,
is too small to account,
in terms of the Sakharov mechanism [1], for the baryon asymmetry of the universe [2].
In this context, a study has been devoted to the possibility of explaining CP violation
in the minimal supersymmetric standard model (MSSM) [3].
The supersymmetry (SUSY) itself is widely regarded as an essential ingredient
for the new physics beyond the SM [4].
Supersymmetrically-extended standard models may have various sources of complex phases,
either explicitly or spontaneously.
These complex phases may arise from the Higgs sector during  the spontaneous symmetry breaking or
from radiative corrections at higher level.
Also, the soft SUSY-breaking terms, which are required as essential ingredients
to construct the full Lagrangian density for practical analysis of the supersymmetric models,
might be complex.

The minimal supersymmetric standard model, which has no Higgs singlet
and, thus, is the simplest version of supersymmetric standard model,
has been investigated exhaustively with respect to various aspects of phenomenology [5],
and the possibility of the CP violation [6].
It has been noticed that it can accommodate CP violation at the one-loop level,
as complex phases in the soft SUSY-breaking terms may induce explicit CP mixings
between the scalar and the pseudoscalar Higgs bosons in the neutral Higgs sector [7].

The Higgs sector of the MSSM may be enlarged by adding a Higgs singlet.
This model is known as the next-to-MSSM, and might be referred to as the (M+1)SSM.
One of the motivations for the enlargement is to solve dynamically the so-called dimensional $\mu$-parameter problem
in the MSSM by means of the vacuum expectation value (VEV) of the Higgs singlet.
The (M+1)SSM, including the CP phenomenology,
has also been investigated by many authors [8-10].
At the tree level, the (M+1)SSM may contain at most one nontrivial CP phase after redefining the Higgs fields.
At the one-loop level, it is observed that there are noticeable effects of explicit CP violation
in the (M+1)SSM on the neutral and charged Higgs boson masses,
when radiative corrections due to relevant particles and their superpartners are taken into account [10].

A further extension of the MSSM with two additional Higgs singlets is also possible.
This model might be referred to as the (M+2)SSM.
Actually, in the literature, some attempts have been reported
to study the viability of spontaneous CP violation
in the Higgs sector of the (M+2)SSM [11].
It has been found that, at the tree level, the spontaneous CP violation might occur in the Higgs sector of the (M+2)SSM.

In principle, at the weak scale, the supersymmetrically-extended standard model,
which has at least two Higgs doublets, may embrace additional Higgs multiplets.
In other words, from the theoretical point of view, one may add an arbitrary number of Higgs singlets to the MSSM.
Let us denote the MSSM with $n$ additional Higgs singlets as the (M+$n$)SSM.
In order for the (M+$n$)SSM to be theoretically interesting and consistent with experimental constraints,
its phenomenological implications should be studied.
With respect to the Higgs phenomenology, it should satisfy two important ingredients
for the no-lose theorem to discover a neutral Higgs boson at the Linear Collider (LC),
the future $e^+e^-$ linear collider.
One ingredient is that the lightest neutral Higgs boson should have a mass controlled
by the electroweak scale, and the other is that at least one of the couplings of the Higgs bosons
to a pair of $Z$ boson is as large as that of the SM.

In Ref. 12, Espinosa and Quir\'{o}s investigated the Higgs sector of the (M+$n$)SSM, under the assumption of CP conservation.
They calculated the upper bound on the lightest scalar Higgs boson mass in the (M+$n$)SSM by using the renormalization group approach.
According to their calculations, the lightest scalar Higgs boson mass of the (M+$n$)SSM
is smaller than 145 GeV,
if the gauge couplings in the theory are assumed to remain perturbative
up to a very large energy, which they take as the Grand Unified Theory scale $\Lambda = 10^{16}$ GeV.
The upper bound on the lightest scalar Higgs boson mass strongly depends on the scale $\Lambda$.
In fact, they found that the upper bound on the lightest scalar Higgs boson mass increase
from 145 GeV to 400 GeV,
as $\Lambda$ changed from $10^{16}$ GeV down to $10^{4}$ GeV.

We are motivated by the previous works of Ref. 12, where CP symmetry is conserved in the Higgs sector of the (M+$n$)SSM.
In this article, we are interested in the possibility of CP violation in the Higgs sector of the (M+$n$)SSM.
We would like to study whether the (M+$n$)SSM may accommodate complex phases in its Higgs sector
for CP violation,
while examining if the two above-mentioned phenomenological ingredients are, indeed, satisfied by it.
An absolute upper bound on the lightest neutral Higgs boson mass will be obtained, and
the possibility of discovering the lightest neutral Higgs boson at the LC will be discussed.

The purpose of this article is not only to satisfy our theoretical curiosity about the (M+$n$)SSM
but also to establish the phenomenological plausibility of the (M+$n$)SSM within the context of CP violation.
In this article, we study CP violation in the Higgs sector of the (M+$n$)SSM.
By taking the vacuum expectation values (VEVs) of the neutral Higgs fields
and the relevant parameters in the tree-level Higgs potential as real,
we assume that at the tree level, the Higgs potential of the (M+$n$)SSM has no complex phase.
CP symmetry may then be violated at the one-loop level in an explicit way as radiative corrections are taken into account.
For radiative corrections, we consider only the contributions from the third generation of quarks and squarks.

We obtain the upper bound on the lightest neutral Higgs boson mass in the (M+$n$)SSM for given $n$
by allowing the relevant parameters to vary within reasonable ranges;
The mass depends evidently on $n$: The mass is found to increase as $n$ grows to 10, but does not exceed 220 GeV for larger $n$.
This maximum value of the upper bound is constrained by the condition that the squared masses of squarks be positive.
Thus, production of the lightest neutral Higgs boson in the (M+$n$)SSM is kinematically allowed at the LC.
We, furthermore, investigate prospects for discovering the neutral Higgs bosons of the (M+$n$)SSM at the LC.
The minimum cross section for producing any one of the neutral Higgs bosons of the (M+$n$)SSM in $e^+e^-$ collisions
via the Higgsstrahlung process is found to decrease as $n$ increases.

This article is organized as follows:
In the next section, we describe the Higgs potential of the (M+$n$)SSM and
calculate the masses of the neutral Higgs bosons in the explicit CP violation scenario at the one-loop level.
We then calculate the production cross section of the neutral Higgs bosons via the Higgsstrahlung process at the LC.
Conclusions are presented in the last section.

\section{The Higgs sector}

The Higgs sector of the (M+$n$)SSM contains the two Higgs doublet superfields
${\cal H}_1^T = ({\cal H}_1^0, {\cal H}^-)$ and ${\cal H}_2^T = ({\cal H}^+, {\cal H}_2^0)$ of the MSSM
and an additional $n$ neutral Higgs singlet superfields ${\cal N}_l$ ($l$ = 1 to $n$).
In terms of these Higgs superfields, the most general form of the superpotential of the (M+$n$)SSM may be written as
\[
        {\cal W}
    = h_b {\cal Q} {\cal H}_1 b_R^c + h_t {\cal Q} {\cal H}_2 t_R^c
    + \sum_{i = 1}^n \lambda_i {\cal H}_1^T \epsilon {\cal H}_2 {\cal N}_i - \sum_{i, j, l = 1}^n {k_{i j l} \over 3} {\cal N}_i {\cal N}_j {\cal N}_l  \ ,
\]
where $\lambda_i$ and $k_{i j l}$ ($i, j, l$ = 1 to n) are dimensionless coupling constants,
and  $\epsilon$ is the usual antisymmetric 2 $\times$ 2 matrix with $\epsilon_{12} = - \epsilon_{21}$ = 1.
Here, for simplicity, the Yukawa couplings for the third generation are taken into account:
The SU(2) doublet superfield ${\cal Q}^T$ = ($t_L, b_L$) consists of the left-handed quark superfields,
the SU(2) singlet superfields $t_R^c$ and $b_R^c$ are the charge conjugate of the right-handed quark superfields
of the third generation,
and $h_t$ and $h_b$ are the top and the bottom Yukawa coupling constants, respectively.

At the tree level, the relevant Higgs potential of the (M+$n$)SSM consists of $D$ terms, $F$ terms, and soft terms as
\begin{equation}
            V^0 =  V_D + V_F + V_S  \ ,
\end{equation}
with
\begin{eqnarray}
    V_D & = &
    {g_2^2\over 8} (H_1^\dagger\hat\sigma H_1 + H_2^\dagger\hat\sigma H_2)^2
            + {g_1^2\over 8}(|H_2|^2-|H_1|^2)^2 \ , \cr
    V_F & = &
    \bigg |\sum_{i = 1}^n \lambda_i N_i \bigg |^2 (|H_1|^2+|H_2|^2) \cr
        & &\mbox{} + \sum_{i = 1}^n \bigg | \lambda_i H_1^T \epsilon H_2
        - \sum_{i, j = 1}^n (k_{i j l} + k_{i l j} + k_{l i j} + k_{l j i} + k_{j i l} + k_{j l i}) {N_i N_j \over 6} \bigg |^2  \ ,  \cr
    V_S &=&
    m_{H_1}^2|H_1|^2 + m_{H_2}^2 |H_2|^2 + \sum_{i = 1}^n m_{N_i}^2 |N_i|^2
        - \bigg (\sum_{i = 1}^n \lambda_i A_{\lambda_i} H_1^T \epsilon H_2 N_i + {\rm H.c.} \bigg )  \cr
        & &\mbox{} - \left (\sum_{i, j, l = 1}^n {k_{i j l} \over 3} A_{k_{i j l}} N_i N_j N_l + {\rm H.c.} \right ) \ ,
\end{eqnarray}
where $g_1$ and $g_2$ are the U(1) and the SU(2) gauge coupling constants, respectively,
$\hat\sigma \equiv (\sigma^1,\sigma^2,\sigma^3)$ are the Pauli matrices,
and $H_1^T = (H_1^0, H^-)$, $H_2^T = (H^+, H_2^0)$, and $N_l$ ($l$ = 1 to $n$) are,
respectively, Higgs doublet fields and Higgs singlet fields.
Note that  the soft SUSY breaking masses $m_{H_1}$, $m_{H_2}$, and $m_{N_i}$ ($i$  = 1 to $n$) are
introduced in $V_S$, where $A_{\lambda_i}$ and $A_{k_{i j l}}$ ($i, j, l$  = 1 to $n$) are
the trilinear soft SUSY breaking parameters with a dimension of mass.
We assume that the parameters in $V^0$ are real.
Thus, the CP symmetry is assumed to be conserved at the tree level.

The total number of degrees of freedom in the Higgs sector of the (M+$n$)SSM
from two Higgs doublets and $n$ Higgs singlets is 2$n$+8.
After the spontaneous breaking of electroweak symmetry, three of them are gauged away,
becoming a neutral Goldstone boson and a pair of charged ones.
The remaining 2$n$+5 degrees of freedom correspond to 2$n$+3 neutral Higgs bosons and a pair of charged Higgs bosons.
In terms of the physical Higgs fields, after the spontaneous breaking of electroweak symmetry,
the Higgs doublets and the $n$ Higgs singlets may be expressed as
\begin{eqnarray}
\begin{array}{lll}
        H_1 & = & \left ( \begin{array}{c}
          v_1 + h_1 + i \sin \beta h_3   \cr
          \sin \beta C^{+ *}
  \end{array} \right )  \ ,  \cr
        H_2 & = & \left ( \begin{array}{c}
          \cos \beta C^+           \cr
          v_2 + h_2 + i \cos \beta h_3
  \end{array} \right )   \ ,  \cr
        N_j & = & \left ( \begin{array}{c}
          x_j + h_{(2j+2)} + i h_{(2j+3)}
  \end{array} \right ) \ ,   \ j = 1 \ {\rm to} \  n  \ ,
\end{array}
\end{eqnarray}
where $h_l$ ($l$ = 1 to 2$n$+3) are the neutral Higgs fields, $C^+$ is the charged Higgs field,
and $v_1$, $v_2$, $x_j$ ($j$ = 1 to $n$) are
the VEVs, respectively, of $H_1$, $H_2$, $N_j$ ($j$ = 1 to $n$).
We assume that the VEVs are real.
With the above form for the Higgs doublets, one may take the constraints $m_W^2 = g_2^2 v^2 /2$ and
$m_Z^2 = (g_1^2 + g_2^2) v^2 /2$ for the electroweak gauge bosons with $v = \sqrt{v_1^2 + v_2^2}$ = 175 GeV.
Note that $\tan \beta$ is given by the ratio of the VEVs of the two Higgs doublets, $v_1/v_2$.

Now, we consider higher-order corrections.
The Higgs sector of supersymmetric models at the tree level is well observed to be
significantly affected by radiative corrections.
The total Higgs potential at the one-loop level may be written as
\[
        V = V^0 +V^1  \  ,
\]
where $V^1$ represents the one-loop corrections.
The effective potential method allows us to obtain $V^1$ as [13]
\[
    V^1
    = \frac{3}{32\pi^2} \left \{ {\cal M}_{\tilde{q_l}}^4
    \left (\log {{\cal M}_{\tilde{q_l}}^2 \over \Lambda^2} - {3\over 2} \right )
    - 2 {\cal M}_q^4 \left (\log {{\cal M}_q^2 \over \Lambda^2}
    - {3\over 2} \right ) \right \}  \ ,
\]
where ${\cal M}_{{\tilde q}_l}^2$ ($l$ = 1 to 4) and ${\cal M}_q^2$ ($q$ = $b, t$) are,
respectively, the field-dependent squark and quark masses for the third generation,
and $\Lambda$ is the renormalization scale in the modified minimal subtraction $(\overline {\rm MS})$ scheme.
We consider here the radiative corrections due to the loops of quarks and squarks of only the third generation.

Note that the squark terms are positive whereas the quark terms are negative in $V^1$.
Because of the opposite sign between the squark and the quark terms,
they would exactly cancel out if the masses of the quarks and the squarks were equal.
Then, there would be no contribution to $V^1$.
However, this is not the case because no squarks with masses comparable to quarks have been observed.
Thus, we assume that the quarks and the squarks of the third generation have different masses,
so the radiative corrections due to them should persist.

The masses of the stop and sbottom quarks are obtained as
\begin{eqnarray}
        m^2_{{\tilde t}_1}, \ m^2_{{\tilde t}_2}
    & = & m_t^2 + {1 \over 2} (m_Q^2 + m_T^2) \mp \sqrt{X_{\tilde t} }    \ , \cr
        m^2_{{\tilde b}_1}, \ m^2_{{\tilde b}_2}
    & = & m_b^2 + {1 \over 2} (m_Q^2 + m_B^2) \mp \sqrt{X_{\tilde b} }  \ ,
\end{eqnarray}
where $m_Q$, $m_T$, and $m_B$ are the soft SUSY-breaking masses,
and $m_t = h_t v_2$ and $m_b = h_b v_1$ are,
respectively,
the top quark mass and the bottom quark mass after the electroweak symmetry breakdown.
The mixing between the two stop quarks is represented by $X_{\tilde t}$, and
the mixing between the two sbottom quarks by $X_{\tilde b}$.
The two stop quarks or the two sbottom quarks will be degenerate in mass if $X_{\tilde t}$ or $X_{\tilde b}$ vanishes.
We assume that they are not degenerate.
They are, respectively, given as
\begin{eqnarray}
    X_{\tilde t} & = &
    {1 \over 4} (m_Q^2 - m_T^2)^2
        + m_t^2 \{A_t^2  + (\sum_{j = 1}^n \lambda_j^2 x_j^2) \cot^2 \beta
    + 2 A_t (\sum_{j = 1}^n \lambda_j x_j) \cot \beta \cos \phi_t \} \ , \cr
    X_{\tilde b} & = &
    {1 \over 4} (m_Q^2 - m_B^2)^2
        + m_b^2 \{A_b^2 + (\sum_{j = 1}^n \lambda_j^2 x_j^2) \tan^2 \beta
    + 2 A_b (\sum_{j = 1}^n \lambda_j x_j) \tan \beta \cos \phi_b \} \ ,
\end{eqnarray}
where $A_t$ and $A_b$ are the trilinear soft SUSY breaking parameters with a dimension of mass.

We assume that $A_t$ and $A_b$ are complex, and we denote their phases by $\phi_t$ and $\phi_b$, respectively.
If $A_t$ and $A_b$ are real, $\phi_t$ and $\phi_b$ will be absent in the above expressions.
Their presence implies CP violation at the one-loop level.
CP symmetry is assumed to be conserved at the tree level, as the VEVs of the neutral Higgs fields
and the relevant parameters in $V^0$ are taken as real.
Now, CP symmetry is allowed to be broken explicitly at the one-loop level
in the Higgs sector of the (M+$n$)SSM by complex $A_t$ and $A_b$.

If CP symmetry is conserved at the tree level and is broken explicitly at the one-loop level in the (M+$n$)SSM,
there are non-trivial CP-odd tadpole minimum conditions for the neutral imaginary parts
of $H_1$, $H_2$, and $N_j$ ($j$ = 1 to $n$), namely, $h_3$ and $h_{(2j+3)}$ ($j$ = 1 to $n$).
Thus, the number of CP-odd tadpole minimum conditions in the (M+n)SSM is $n+1$.
They read
\[
    0 =
    {3 m_t^2 A_t (\sum_{j = 1}^n \lambda_j x_j) \sin \phi_t \over 16 \pi^2 v^2 \sin^2 \beta}
    f(m_{{\tilde t}_1}^2, \ m_{{\tilde t}_2}^2)
    + {3 m_b^2 A_b (\sum_{j = 1}^n \lambda_j x_j) \sin \phi_b \over 16 \pi^2 v^2 \cos^2 \beta}
    f(m_{{\tilde b}_1}^2, \ m_{{\tilde b}_2}^2)
\]
for $h_3$ and
\[
    0 =
    {3 m_t^2 A_t \lambda_j x_j \sin \phi_t \over 16 \pi^2 v \sin^2 \beta}
    f(m_{{\tilde t}_1}^2, \ m_{{\tilde t}_2}^2)
    + {3 m_b^2 A_b \lambda_j x_j \sin \phi_b \over 16 \pi^2 v \cos^2 \beta}
    f(m_{{\tilde b}_1}^2, \ m_{{\tilde b}_2}^2)
\]
for $h_{2j+3}$ ($j$ = 1 to $n$), where the scale-dependent function $f$ is given by
\[
        f(m_x^2, \  m_y^2)
            = {1 \over (m_y^2-m_x^2)} \left \{m_x^2 \log  {m_x^2 \over
            \Lambda^2} -m_y^2 \log {m_y^2 \over \Lambda^2} \right \} + 1 \ .
\]
As can be seen from the structure of the above equations, the $n$+1 CP-odd tadpole minimum conditions may be
reduced into just one equation, namely,
\[
    0 =
    {m_t^2 A_t \sin \phi_t \over \sin^2 \beta}
    f(m_{{\tilde t}_1}^2, \ m_{{\tilde t}_2}^2)
    + {m_b^2 A_b \sin \phi_b \over \cos^2 \beta}
    f(m_{{\tilde b}_1}^2, \ m_{{\tilde b}_2}^2) \ .
\]
Further, the soft SUSY-breaking masses  $m_{H_1}^2$, $m_{H_2}^2$, and $m_{N_j}^2$ ($j$ = 1 to $n$) are
eliminated by using the $n$+2 minimum conditions.

The squared masses of the 2$n$+3 neutral Higgs bosons, $m^2_{h_l}$ ($l$ = 1 to 2$n$+3),
are given by the eigenvalues of a mass matrix.
These eigenvalues are obtained by the second derivatives
of the Higgs potential with respect to the neutral Higgs fields $h_l$ ($l$ = 1 to 2$n$+3),
where the minimum conditions are employed.
The symmetric (2$n$+3)$\times$ (2$n$+3) mass matrix for the neutral Higgs bosons, $M_{ij}^2$,
may be expressed in the basis of ($h_1$, $h_2$, $\cdots$, $h_{(2n+3)}$).
It is understood that, after diagonalizing the mass matrix, we sort the eigenvalues
in increasing order by rearranging the neutral Higgs fields
such that  the (1,1)-element of the diagonalized mass matrix is $m_{h_1}^2$,
the (2,2)-one is $m_{h_2}^2$, and so on.
Thus, we implicitly assume hereafter that $h_1$ is not the neutral real component
of the $H_1$ Higgs field, but the lightest neutral Higgs boson, $h_2$ is not that of $H_2$
but the second lightest neutral Higgs boson, and so on.

Let us concentrate on the upper-left 2 $\times$ 2 sub-matrix of $M_{ij}^2$.
It is mathematically known that the upper bound on $m_{h_1}^2$ can be obtained
from the upper-left 2 $\times$ 2 sub-matrix of $M_{ij}^2$, regardless of the remaining matrix elements.
The upper bound on $m_{h_1}^2$ is then obtained as
\begin{eqnarray}
    m_{{h_1}, \ {\rm max}}^2  = &&
    m_Z^2 \cos^2 2 \beta + v^2 (\sum_{j = 1}^n \lambda_j)^2 \sin^2 2 \beta \cr
& &\mbox{} + {3 m_t^4 \over 8 \pi^2 v^2}
    {\{ (\sum_{j = 1}^n \lambda_j) (\sum_{j = 1}^n x_j) \cot \beta \Delta_{{\tilde t}_1} + A_t \Delta_{{\tilde t}_2} \}^2
    \over (m_{{\tilde t}_2}^2 - m_{{\tilde t}_1}^2)^2}
    g(m_{{\tilde t}_1}^2, \ m_{{\tilde t}_2}^2) \cr
&  & \mbox{} + {3 m_t^4 \over 4 \pi^2 v^2}
    {\{(\sum_{j =1}^n \lambda_j) (\sum_{j = 1}^n x_j) \cot \beta \Delta_{{\tilde t}_1} + A_t \Delta_{{\tilde t}_2} \}
    \over (m_{{\tilde t}_2}^2 - m_{{\tilde t}_1}^2)}
    \log \left({m_{{\tilde t}_2}^2 \over m_{{\tilde t}_1}^2}\right) \cr
&  &\mbox{} + {3 m_t^4 \over 8 \pi^2 v^2}
    \log \left ({m_{{\tilde t}_1}^2 m_{{\tilde t}_2}^2 \over m_t^4} \right ) \cr
& &\mbox{} + {3 m_b^4 \over 8 \pi^2 v^2}
    {\{(\sum_{j = 1}^n \lambda_j) (\sum_{j = 1}^n x_j) \tan \beta \Delta_{{\tilde b}_1} + A_b \Delta_{{\tilde b}_2} \}^2
    \over (m_{{\tilde b}_2}^2 - m_{{\tilde b}_1}^2)^2}
    g(m_{{\tilde b}_1}^2, \ m_{{\tilde b}_2}^2) \cr
& &\mbox{} + {3 m_b^4 \over 4 \pi^2 v^2}
    {\{(\sum_{j = 1}^n \lambda_j) (\sum_{j = 1}^n x_j) \tan \beta \Delta_{{\tilde b}_1} + A_b \Delta_{{\tilde b}_2} \}
    \over (m_{{\tilde b}_2}^2 - m_{{\tilde b}_1}^2)}
    \log \left({m_{{\tilde b}_2}^2 \over m_{{\tilde b}_1}^2}\right) \cr
&  &\mbox{} + {3 m_b^4 \over 8 \pi^2 v^2}
    \log \left ({m_{{\tilde b}_1}^2 m_{{\tilde b}_2}^2 \over m_b^4} \right )
\end{eqnarray}
by using the two inequalities
\begin{eqnarray}
    && (\sum_{j = 1}^n \lambda_j)^2  \ge \sum_{j = 1}^n \lambda_j^2 \ , \cr
    && (\sum_{j = 1}^n \lambda_j) (\sum_{j = 1}^n x_j)  \ge \sum_{j = 1}^n \lambda_j x_j \ ,
\nonumber
\end{eqnarray}
where
\begin{eqnarray}
    \Delta_{{\tilde t}_1} & = &
    A_t \cos \phi_t + (\sum_{j = 1}^n \lambda_j) (\sum_{j = 1}^n x_j) \cot \beta  \  , \cr
    \Delta_{{\tilde t}_2} & = &
    A_t + (\sum_{j = 1}^n \lambda_j) (\sum_{j = 1}^n x_j) \cot \beta \cos \phi_t  \ , \cr
    \Delta_{{\tilde b}_1} & = &
    A_b + (\sum_{j = 1}^n \lambda_j) (\sum_{j = 1}^n x_j) \tan \beta \cos \phi_b  \ , \cr
    \Delta_{{\tilde b}_2} & = &
    A_b \cos \phi_b + (\sum_{j = 1}^n \lambda_j) (\sum_{j = 1}^n x_j) \tan \beta   \ ,
\end{eqnarray}
with the scale-independent function $g$ given by
\[
    g(m_x^2, m_y^2) = {m_y^2 + m_x^2 \over m_x^2 - m_y^2}  \log {m_y^2 \over m_x^2} + 2  \ .
\]
It can be observed that $m_{h_1, {\rm max}}^2$ depends on $\lambda_j$ ($j$ = 1 to $n$).
As $\lambda_l$ increases, $m_{h_1, {\rm max}}^2$ increases.
It is worthwhile noticing that, if we replace $\sum_{j = 1}^n \lambda_j$
by $\lambda$ and $\sum_{j = 1}^n x_j$ by $x$, the above expression for $m_{{h_1}, \ {\rm max}}^2$ looks
exactly the same as the corresponding formula for the upper bound on the lightest Higgs boson mass
in the (M+1)SSM, where $\lambda$ and $x$ are, respectively, the dimensionless coupling constant
and the VEV of the Higgs singlet in the (M+1)SSM.

Now, in order to obtain the numerical value for $m_{{h_1}, \ {\rm max}}^2$,
one has to take the values of $\lambda_j$ and $x_j$ ($j$ = 1 to $n$),
as well as other relevant parameter values.
For simplicity, we set  $A_t = A_b$ and $m_Q = m_T = m_B$.
We choose a region of parameter space defined by $2 \le \tan \beta \le 40$,
$0 < \lambda_j \le 0.7$, $0 < x_j, m_Q, A_t \le$ 1000 GeV ($j$ = 1 to $n$),
and $\pi/100 \le \phi_t, \phi_b \le 99 \pi/100$.
We examine randomly $10^6$ points in the defined region of the parameter space.
For each point, we check if  $m_{{\tilde q}_1} > m_t = 175$ GeV and
if $m_{{\tilde q}_2} \neq m_{{\tilde q}_1}$ $(q = t, b)$.
The first condition is physically justified because it makes $m_{{h_1}, \ {\rm max}}^2$
more restrictive; and the second condition is required for numerical stability.
If these two conditions are satisfied, we then calculate $m_{{h_1}, \ {\rm max}}^2$.
In this way, we obtain the upper bound on the mass of the lightest neutral Higgs boson
in the (M+$n$)SSM for given $n$, which is an absolute upper bound in the sense that
we examine the whole meaningful parameter space.

Figure 1 shows $m_{{h_1}, \ {\rm max}}^2$ plotted against $n$, the number of Higgs singlets in the (M+$n$)SSM.
As $n$ increases, $m_{{h_1}, \ {\rm max}}^2$ increases from 148 GeV for $n=1$.
However, notice that it does not grow indefinitely.
We find that it reaches 220 GeV for $n$= 15 and then very gradually decreases for $n > 15$.
Thus, the universal upper bound on $m_{h_1}$ in the (M+$n$)SSM at the one-loop level
for arbitrary number of Higgs singlets might be said to be 220 GeV.

The upper bound on $m_{h_1}$ is actually set by the condition
that the squared masses of the scalar quarks should be positive.
From the expressions for the squark masses shown in Eq. 4,
it is clear that $\sqrt{X_{\tilde q}}$ ($q$ = $t$ or $b$) might be restricted from above
in order for the squared masses of the scalar quarks to be positive.
Since  the maximum value of $m_Q$ $(= m_T = m_B)$ is on a SUSY-breaking scale ($\sim$ 1000 GeV),
$\sqrt{X_{\tilde q}}$ ($q$ = $t$ or $b$) must be smaller than (1000 GeV)$^2$.
Otherwise, the electroweak symmetry breaking will not occur.
Moreover, we have assumed that the lighter scalar quark masses are larger than the top quark mass.

We note that in the (M+$n$)SSM, one can impose two additional constraints on the parameter space.
For any SUSY models in general, the heavier scalar quark masses must be smaller
than the SUSY-breaking scale ($\sim$ 1000 GeV).
That is, an absolute difference between the mass of any ordinary particle and
the mass of its corresponding superpartner must be controlled by the SUSY-breaking scale.
In particular, in SUSY models with Higgs singlets, the value of $(\sum_{j = 1}^n \lambda_j )(\sum_{j = 1}^n x_j)$
is limited by the electroweak scale $m_{\rm EW} \sim$ 300 GeV.
If these two conditions are further imposed on the parameter space,
 $m_{{h_1}, \ {\rm max}}^2$ should be lowered from 220 GeV.
In our calculations, these two constraints are not imposed.
Thus, our result is a more conservative value.

For the other neutral Higgs bosons, one can obtain an upper bound
on $m^2_{h_l}$ ($l$ = 2 to 2$n$+3) as functions of $m^2_{{h_1}{\rm ,max}}$ and $m^2_{h_1}$.
Let $O$ be the (2n+3)$\times$(2n+3) orthogonal matrix
that diagonalizes the neutral Higgs boson mass matrix.
We now introduce
\[
    R_l = O_{l1} \cos \beta + O_{l2} \sin \beta
\]
($l$ = 1 to $2n+3$).
It is clear that $0 \le R_l^2 \le 1$ and that a kind of orthogonality condition,
namely, $\sum_{l = 1}^{(2n+3)} R_l^2 = 1$, is satisfied.
The $R_l$ represent collectively the whole parameter space of the (M+$n$)SSM.
Thus, it is more convenient to consider them than examining the parameter space itself
because a particular set of $R_l$ ($l$ = 1 to $2n+3$) corresponds to a particular set
of parameter values of the (M+$n$)SSM.
In terms of $R_l$ ($l$ = 1 to $2n+3$), the upper bounds on the squared masses
for the rest of neutral Higgs bosons are known to be given by
\[
        m_{h_l{\rm ,max}}^2
        = { m_{h_1{\rm ,max}}^2 - (\sum_{j = 1}^{l -1} R_j^2) m_{h_1}^2 \over 1 - (\sum_{j = 1}^{l -1} R_j^2) }  \ ,
\]
for $l$ = 2  to 2$n$+3.

In the (M+$n$)SSM, we expect that $ZZh_l$ ($l$ = 1 to $2n+3$), the coupling of a neutral Higgs boson
to a pair of neutral gauge bosons, to decrease if $n$ increases
because the dimension of the neutral Higgs boson mass matrix increases
by two for each additional Higgs singlet.
Thus, for example, the production cross section of a neutral Higgs boson
via the Higgsstrahlung process $e^+e^- \rightarrow Z h_l$ in $e^+e^-$ collisions decreases
if  $n$ increases.
In fact,  the production cross section
via the Higgsstrahlung process gets smaller if more Higgs singlets are added
to the model (a large number of singlets is predicted in some string models) or
if there is CP violation with a mixing of scalar and pseudoscalar Higgs bosons [14].

Now, let us consider the possibility of discovering any one of the neutral Higgs bosons
in the (M+$n$)SSM at the LC.
We calculate the production cross sections of neutral Higgs bosons
via the Higgsstrahlung process in $e^+e^-$ collisions.
For the center of mass energy of the future LC, we take  $\sqrt{s}$ = 500 GeV, 1000 GeV, and 2000 GeV.

Whether or not any one of $h_l$ ($l$ = 1 to $2n+3$) might be produced
via the Higgsstrahlung process with real $Z$ is determined
by the threshold energy of $E_{\rm T} = m_Z + h_l$.
For the LC with $\sqrt{s}$ = 500, 1000, and 2000 GeV,
$h_1$ production is kinematically allowed because $m_{h_1} < 220$ GeV.
If the production cross section of $h_1$ via the Higgsstrahlung process is small,
one should then examine whether any of the other $h_l$ ($l$ = 2 to 2n+3) has
a production cross section large enough to be detected at the LC.

Let $\sigma_{\rm SM}(m)$ be the production cross section for the SM Higgs boson of mass $m$
via the Higgsstrahlung process at a given center of mass energy.
Then, in the (M+$n$)SSM, the production cross section $\sigma_l$ ($l$ = 1 to 2$n$+3)
for the $l$-th neutral Higgs boson via the Higgsstrahlung process at a given center of mass energy is given by
\[
        \sigma_l (m_{h_l}) = \sigma_{\rm SM} (m_{h_l}) R_l^2  \ .
\]
In particular, the production cross section for the heaviest neutral Higgs boson may be written as
\[
        \sigma_{\rm (2n+3)} (m_{h_{\rm (2n + 3)}})
    = \sigma_{\rm SM} (m_{h_{\rm (2n+3)}}) \left (1 - \sum_{l = 1}^{\rm (2n+2)} R_l^2 \right ) \ .
\]
Since the production cross section via the Higgsstrahlung process decreases
as the neutral Higgs boson mass increases, we have for any of them
\[
    \sigma(m) \ge \sigma(m_{\rm max}) = \sigma_{\rm min}  \ .
\]
Hence, for given $m_{h_l}$ and $R_l$ ($l$ = 1 to 2$n$+3), $\sigma_l (m_{h_l})$ satisfy
\begin{eqnarray}
        && \sigma_1 = \sigma_1 (R_1, m_{h_1})  \ge \sigma_1 (m_{h_1, {\rm max}})  = \sigma_{1, {\rm min}}  \ , \cr
    && \sigma_2 = \sigma_2 (R_1, R_2, m_{h_2})  \ge  \sigma_2 (m_{h_2, {\rm max}})  = \sigma_{2, {\rm min}} \ ,
\nonumber
\end{eqnarray}
and so on.
The minimum values, $\sigma_{l, {\rm min}}$ ($l$ = 1 to 2$n$+3), depend
on $m_{h_1, {\rm max}}$ and $R_l$ ($l$ = 1 to 2$n$+3) such that
\begin{eqnarray}
    && \sigma_{1, {\rm min}}  = \sigma_{1, {\rm min}} (m_{h_1, {\rm max}})   \ , \cr
    && \sigma_{2, {\rm min}}  =  \sigma_2 (m_{h_2, {\rm max}})
                  = \sigma_{2, {\rm min}} (R_1, m_{h_1, {\rm max}})  \ ,  \cr
    && \sigma_{3, {\rm min}}  =  \sigma_3 (m_{h_3, {\rm max}})
                  = \sigma_{3, {\rm min}} (R_1, R_2, m_{h_1, {\rm max}})  \ ,
\end{eqnarray}
and so on, because $m_{h_l{\rm ,max}}^2$ depend on $R_l$ ($l$ = 1 to 2$n$+3).
Examining these 2$n$+3 values $\sigma_{l, {\rm min}}$ ($l$ = 1 to 2$n$+3),
we may determine the possibility of detecting one of the 2$n$+3 Higgs bosons in $e^+e^-$ collisions.

For given $n$, $m_{h_1, {\rm max}}$ is fixed between 148 and 220 GeV.
Then, for each $n$ and for given $R^2_1$ between 0 and 1,
we randomly generate  $R^2_l$ ($l$ = 2 to 2$n$+3) between 0 and 1
with the constraint $\sum_{l = 1}^{(2n+3)} R_l^2 = 1$ in order to calculate
$\sigma_{l, {\rm min}}$ ($l$ = 1 to 2$n$+3).
Actually, this job is equivalent to exploring the whole parameter space of the (M+$n$)SSM.
Among the 2$n$+3 $\sigma_{l, {\rm min}}$ ($l$ = 1 to 2$n$+3) for given $n$,
let us select the largest one and denote it as
\begin{equation}
        \sigma^{\rm max} (R^2_1)    \ .
\end{equation}
This value is the minimum production cross section, above which
at least one of the 2$n$+3 neutral Higgs bosons is produced
via the Higgsstrahlung process in $e^+e^-$ collisions.
If we select, on the other hand, the smallest of  them,
it is the minimum production cross section, above which all of the 2$n$+3 neutral Higgs bosons
are produced via the Higgsstrahlung process in $e^+e^-$ collisions.

We first set $n = 1$, and check the consistency of our analysis with previous studies in the (M+1)SSM.
In the (M+1)SSM, there is only one Higgs singlet and five neutral Higgs bosons.
We plot in Fig. 2 $\sigma^{\rm max} (R^2_1)$ in the (M+1)SSM
with explicit CP violation as a function of $R^2_1$, for $\sqrt{s} = 500$, 1000, and 2000 GeV.
Our results are consistent with previous studies.
For $0 \le R^2_1 \le 1$, that is, for the entire parameter space of the (M+1)SSM,
the $\sigma^{\rm max} (R^2_1)$ seems to be well-behaved.

From Fig. 2, we find that the $\sigma^{\rm max} (R^2_1)$ in (M+1)SSM
at $\sqrt{s} = 500$ GeV is no smaller than 10 fb.
This implies that at least one of the five neutral Higgs bosons
in the (M+1)SSM might be produced in $e^+e^-$ collisions
for a certain set of parameter values of the (M+1)SSM,
with its production cross section being larger than 10 fb.
If the c.m. energy of the $e^+e^-$ system increases,
we find that the minimum value of $\sigma^{\rm max} (R^2_1)$ in (M+1)SSM decreases:
At $\sqrt{s} = 1000$ GeV,  $\sigma^{\rm max} (R^2_1)$ is no smaller than 2.5 fb, and at  $\sqrt{s} = 2000$ GeV,
$\sigma^{\rm max} (R^2_1)$ is no smaller than 0.6 fb.
Therefore, we may expect the integrated luminosity for 50 events of detecting one of
the five neutral Higgs bosons
in the (M+1)SSM to be about 10 fb$^{-1}$ for the LC with $\sqrt{s} = 500$ GeV, about 40 fb$^{-1}$
for the LC with $\sqrt{s} = 1000$ GeV, and about 166 fb$^{-1}$ for the LC with $\sqrt{s} = 2000$ GeV
if we assume that the efficiency of the machine is about 50\%.

Now, let us consider the (M+$n$)SSM for $n > 1$.
One may introduce a dimensionless parameter $\rho$ in order to investigate the size of the CP mixing
between the scalar and the pseudoscalar Higgs bosons.
This parameter is defined in the (M+$n$)SSM by
\begin{equation}
        \rho = {(2n+3)} \left [ \sqrt[2n+3]{R_1^2 R_2^2 \cdots R_{2n+3}^2} \right ]  \ .
\end{equation}
The range of $\rho$ is from 0 to 1 because the $R_l$ ($l$ = 1 to 2$n$+3) satisfy the condition
of $\sum_{l = 1}^{\rm 2n+3} R^2_l = 1$.
If $\rho = 0$, there is no explicit CP violation in the Higgs sector of the (M+$n$)SSM.
On the other hand, if $\rho = 1$, CP symmetry is maximally violated by the mixing
between the scalar and the pseudoscalar Higgs bosons.
The maximal CP violation that leads to $\rho = 1$ takes place when $R_1^2 = R^2_2 = R^2_3 = \cdots = 1/{\rm (2n + 3)}$.
In the case of maximal CP violation, the neutral Higgs boson search is extremely difficult.
Actually, this can be seen in Fig. 2 for (M+1)SSM, where $\sigma^{\rm max} (R^2_1)$ becomes smallest at  $R_1^2 \sim 1/5$.
In the (M+1)SSM where we have five neutral Higgs bosons and $R_1^2 = R^2_2 = \cdots = R^2_5$,
the maximal CP violation will take place when $R_1^2 = R^2_2 = \cdots = R^2_5 = 1/5$.
We deduce from Fig. 2 that the minimum of $\sigma^{\rm max} (R^2_1)$ occurs at $R_1^2 \sim 1/5$ in the (M+1)SSM
when the rest parameters $R^2_j$ ($j$ = 2 to 5) also become incidentally about 0.2 during random generation between 0 and 1.

For each $n > 1$, we repeatly determine the minimum of $\sigma^{\rm max} (R^2_1)$ for $\sqrt{s}$ = 500, 1000, and 2000 GeV.
Our result is shown in Fig. 3, where the minimum of $\sigma^{\rm max} (R^2_1)$ is
plotted against $n$ for $\sqrt{s}$ = 500, 1000, and 2000 GeV.
We see that the minimum of $\sigma^{\rm max} (R^2_1)$ decreases monotonically as $n$ increases:
This is because the coupling coefficient of the Higgs bosons to the $Z$ boson pair,
at the $ZZh_j$ vertex, decreases as $n$ increases.
For example, for $\sqrt{s}$ = 2000 GeV, the minimum of $\sigma^{\rm max} (R^2_1)$ is
as small as about 0.08 fb for $n =$ 15.

In Fig. 4, we plot the integrated luminosity for 50 events of Higgs production
against $n$ in units of fb$^{-1}$ for $\sqrt{s}$ = 500, 1000, and 2000 GeV,
where the efficiency of the $e^+e^-$ collider is assumed to be about 50\%.
For example, for $n =$ 15, the integrated luminosity should be about 1084 ${\rm fb}^{-1}$
for the future LC with $\sqrt{s}$ = 2000 GeV in order to detect 50 events of Higgs production
via the Higgsstrahlung process.

\section{Conclusions}

We study the (M+$n$)SSM, a non-minimal supersymmetric standard model with $n$ Higgs singlets
and examine the possibility of CP violation in the model.
We assume the CP symmetry is intact at the tree level, but is explicitly violated at the one-loop level in the Higgs sector.
CP violation is induced by the mixing between scalar and pseudoscalar Higgs bosons
due to the radiative corrections that are contributed by quarks and squarks of the third generation.
In the presence of CP violation at the one-loop level, we calculate the upper bound
on the mass of the lightest of the 2$n$+3 neutral Higgs bosons and their productions
via the Higgsstrahlung process in $e^+e^-$ collisions at $\sqrt{s}$ = 500, 1000, and 2000 GeV.

The results of our study for $n$ = 1 are consistent with previous studies for the (M+1)SSM.
Our calculations yield an upper bound on the mass of the lightest neutral Higgs boson for $n$ = 1
of about 148 GeV.
The production cross section for at least one of the five neutral Higgs bosons of the (M+1)SSM
via the Higgsstrahlung process in $e^+e^-$ collisions at $\sqrt{s}$ = 500 GeV,
according to our calculations, is no smaller than 10 fb.

We also study the model for other values of $n >1$.
The upper bound on the mass of lightest neutral Higgs boson increases up to 220 GeV for $n$ = 15.
This value of 220 GeV might be a universal upper bound on the mass of the lightest neutral Higgs boson
at the one-loop level in the (M+$n$)SSM for any number of Higgs singlets.
This result suggests that the Higgs sector of the (M+$n$)SSM is regulated,
showing no divergence or uncontrollable behavior for large $n$.

The possibility of discovering at least one of the 2$n$+3 neutral Higgs bosons in $e^+e^-$ collisions
via the Higgsstrahlung process was examined by calculating the minimum of the production cross section.
We find that the minimum of the production cross section decreases monotonically for $\sqrt{s}$ = 500 GeV,
from about 10 fb for $n$ = 1 to about 2.5 fb for $n$ = 15.
In terms of the luminosity, we require higher luminosity for the future LC
if we try to discover at least one of 2$n$+3 neutral Higgs bosons in the (M+$n$)SSM
via the Higgsstrahlung process for larger $n$.
Our calculations suggest that an integrated luminosity of about 10 ${\rm fb}^{-1}$
for $n$ =1 to about 80 ${\rm fb}^{-1}$ for $n$ =15  is required
in order to detect 50 events of Higgs production at a future LC with $\sqrt{s}$ = 500 GeV
via the Higgsstrahlung process.
We anticipate that a future LC with sufficient luminosity will yield experimental clues
that may tell us how many Higgs singlets may be accommodated in the supersymmetric standard model.

\section*{ACKNOWLEDGMENT}
This work was supported by Konkuk University in 2007.



 \newpage
\begin{figure}[t]
\begin{center}
\includegraphics[scale=0.6]{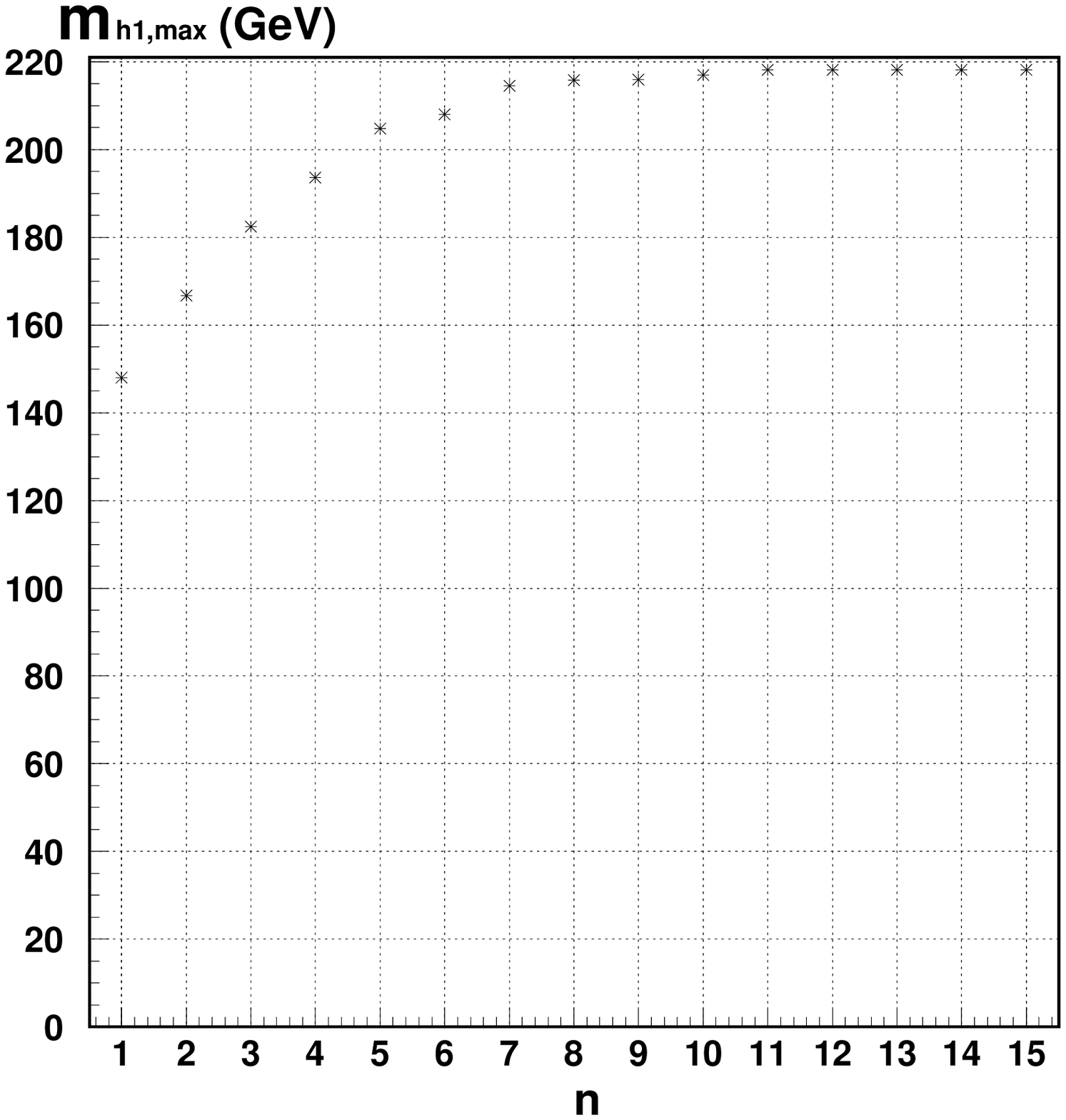}
\caption[plot]{The upper bound on the mass of the lightest neutral Higgs boson against the number of Higgs singlets, $n$.}
\end{center}
\end{figure}

\begin{figure}[t]
\begin{center}
\includegraphics[scale=0.6]{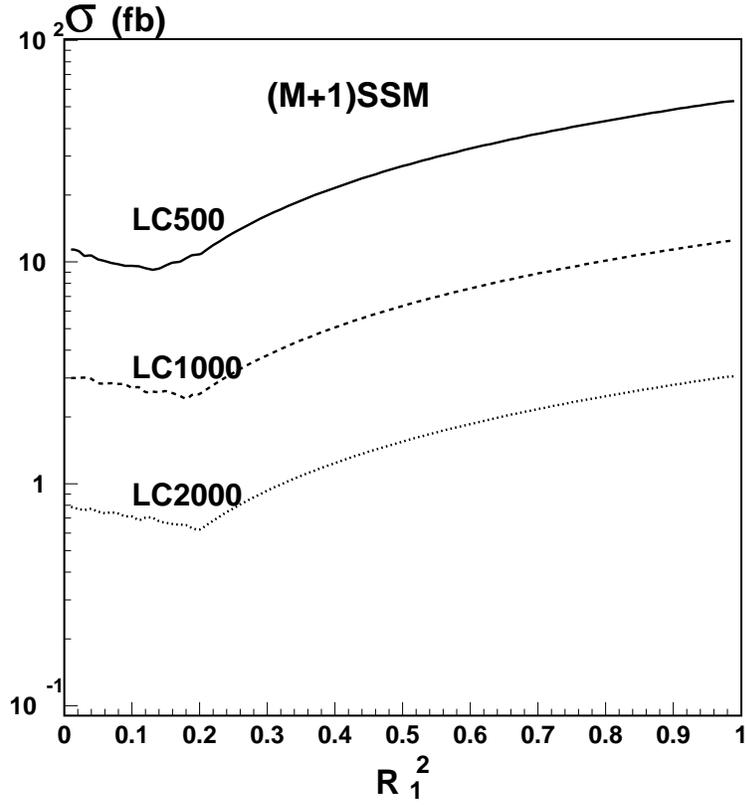}
\caption[plot]{A plot as a function of $R_1^2$ of the production cross section
$\sigma^{\rm max} (R^2_1)$, as defined in the text, for at least one of the five neutral Higgs bosons
in the (M+1)SSM via the Higgsstrahlung process in $e^+e^-$ collisions
with $\sqrt{s}$ = 500, 1000, and 2000 GeV.
The would-be produced neutral Higgs boson is chosen such that the minimum of
its production cross section is larger than the minimum of the production cross section
of any other Higgs bosons in the model.
Note that the curves have minima at $R^2_1 \sim 1/5$.}
\end{center}
\end{figure}

\begin{figure}[t]
\begin{center}
\includegraphics[scale=0.6]{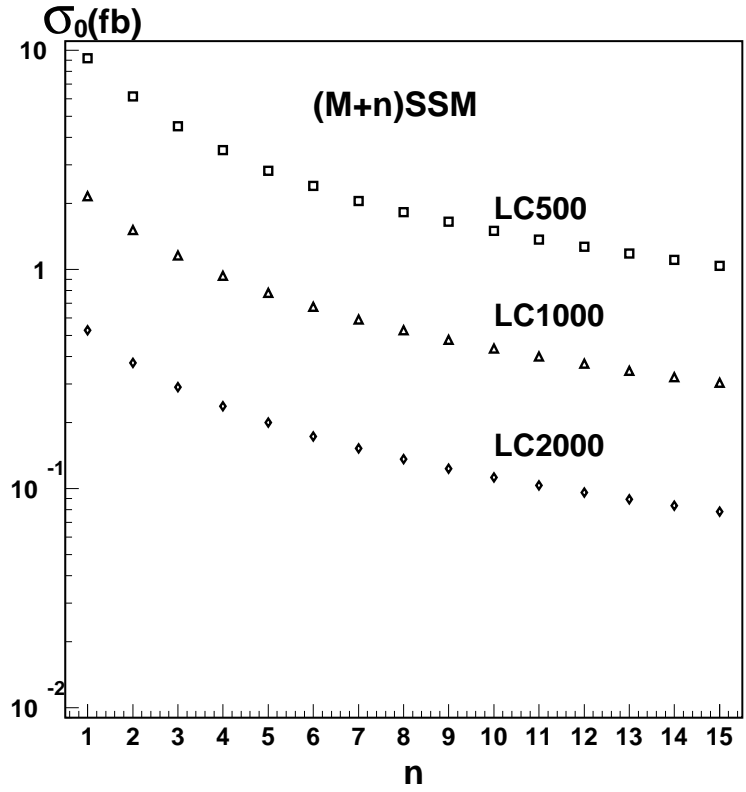}
\caption[plot]{A plot against $n$ of the minimum of $\sigma^{\rm max} (R^2_1)$ at $R^2_1$ = 1/(2$n$+3)
for the future LC with $\sqrt{s}$ = 500, 1000, and 2000 GeV.}
\end{center}
\end{figure}

\begin{figure}[t]
\begin{center}
\includegraphics[scale=0.6]{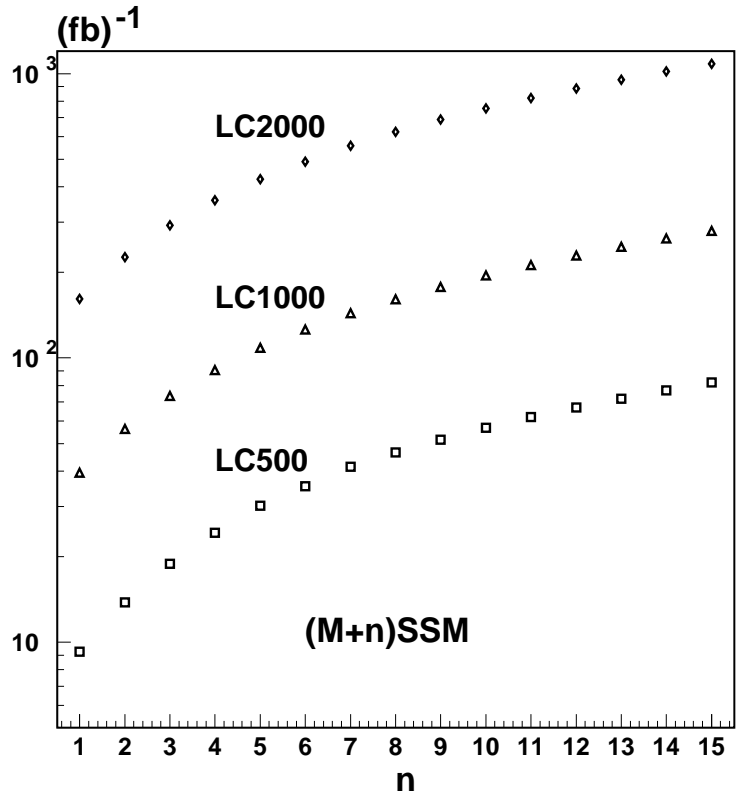}
\caption[plot]{A plot against $n$ of the required integrated luminosity for 50 events of Higgs production
via the Higgsstrahlung process in the future LC with $\sqrt{s}$ = 500, 1000, and 2000 GeV,
where 50\% machine efficiency is assumed.}
\end{center}
\end{figure}

\end{document}